\documentstyle[11pt,jkas2]{article}
\input psfig.tex
\begin{document}

\title{NEW TWISTS IN THE STUDY OF GRAVITY WAVE EMISSION\\
IN SYSTEMS WITH MASSIVE BLACK HOLES }

\author{Sandip K. Chakrabarti}
\address{Tata Institute Of Fundamental Research, Mumbai, 400005, INDIA}

\maketitle

Traditionlly, gravitational wave emission from a coalescing binary
system is computed using point mass approximations (See, e.g., Blanchet
et al. 1995 and references therein.) without considering any accretion 
disk. However, it is believed that in many of the galactic nuclei, 
there are supermassive central black holes, with mass 
$\sim 10^{8-9} M_\odot$ surrounded by accretion disks. These
accretion disks must be necessarily supersonic on and outside the
horizon (Chakrabarti, 1996ab) simply because the radial velocity
(in the corotating frame) has to be the velocity of light on the horizon
while the sound speed must be smaller. However, supersonic
flows are typically sub-Keplerian. Thus, smaller black holes 
and neutron stars (which are on instantaneously Keplerian orbit and 
lose energy and angular momentum through gravitational radiation) 
on their way to coalesce with the central black hole must accrete 
negative angular momentum from the disk. Assuming that the disk becomes highly
sub-Keplerian, the ratio of the rate of lose of angular momentum due 
to such accretion and that due to gravity waves is given by (Chakrabarti,
1996c), $R \sim 1.5 \times 10^{-7} {\rho_{10}}{{T_{10}}^{-3/2}}
x^4 M_8^2$, where $\rho_{10}$, $T_{10}$, $x$ and $M_8$ are the
density in units of $10^{-10}$ g cm$^{-3}$, temperature in units of
$10^{10}$K, binary separation in units of Schwarzschild radius
(of the primary) and $M_8$ is the mass of the primary in units of 
$10^8M_\odot$. At around $x=10$, where, typically, the pressure maximum
occurs in a thick accretion disk, and with $M_8=10=\rho_{10}=T_{10}=1$, 
we have $R=0.015$. Thus, for every one hundred orbits, the companion will
lose $1.5$ orbits due to accretion of negative angular momentum from 
the disk. Here, the frequency of the gravity wave is about $0.008$Hz. 
These waves, together with the deviation from a standard diskless 
two-body coalescence template could be easily detectable by the 
recently proposed gravitational wave detectors, such as LISA and VIRGO. 

As an example, in Fig. 1a, we show the ratio $R_l=l_{disk}/l_{Kep}$,
of the disk and the Keplerian angular momentum distributions
in the case when the flow passes through the inner sonic point at $x=2.3$.
Other parameters are $\gamma=5/3$, $l_{in}=1.7$ and 
the viscosity parameter is:  $\alpha=0.02$. The disk deviates
from a Keplerian disk at $r_{tr}=90$ for these parameters. In Fig.
1b, we show the ratio $R$ as a function of the radial distance 
when the accretion rate is $1000{\dot M}_{Eddington}$. The peak
correction of the loss rate  due to the presence of the
disk is about ten percent which linearly goes down 
with the accretion rate. Fig. 1c shows the number of times the
companion orbits the primary with (solid) or without (dashed)
the disk component as a function of time origin of which
chosen at the instant the disk deviates from a Keplerian disk.

The important point to note is that this result is independent of the
mass of the companion as long as it is very small compared to
the primary. Second, the phase shift from a standard template
(computed without the accretion disk) directly determines the
nature of the sub-Keplerian flow. Third, since sub-Keplerian
flow also act as a Compton cloud (which may be hot or cold
depending on the accretion rate), the fitting of the spectra
in UV/soft X-ray regime with standard models (Chakrabarti \& Titarchuk, 1995)
from the same galaxy would give additional information 
(such as the accretion rate, the optical depth and
the size of the cloud). Together with the phase shift
of the gravitational wave data, one could obtain a complete 
picture of the accreting black hole system at the galactic center. 

\psfig{figure=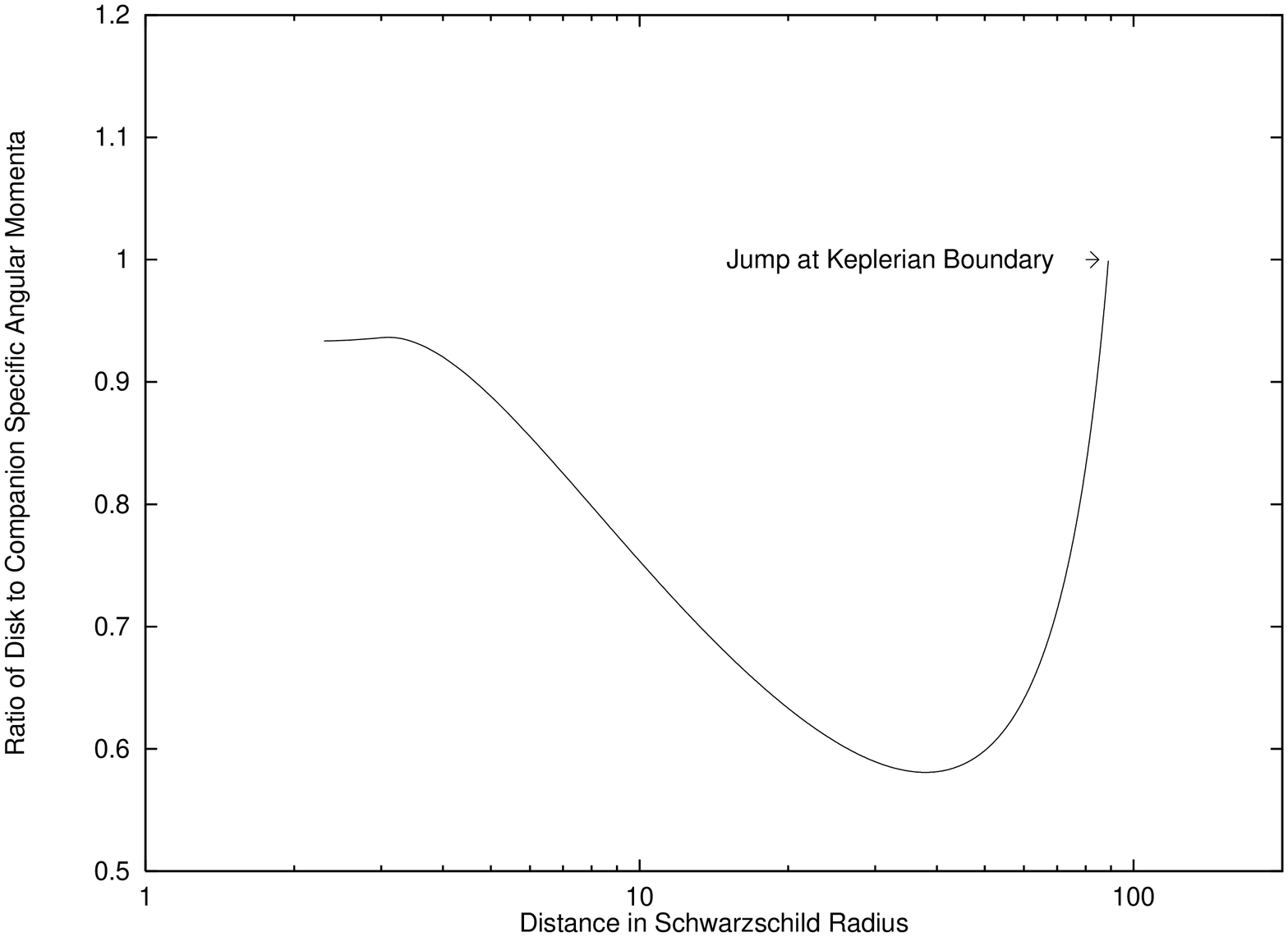,height=7.5truecm,width=7.5truecm,angle=270}
\noindent {\small {\bf Fig. 1a}: Ratio of the disk to Keplerian 
angular momentum distributions after the flow deviates from a 
Keplerian disk at $x=90$.}

\bigskip

\psfig{figure=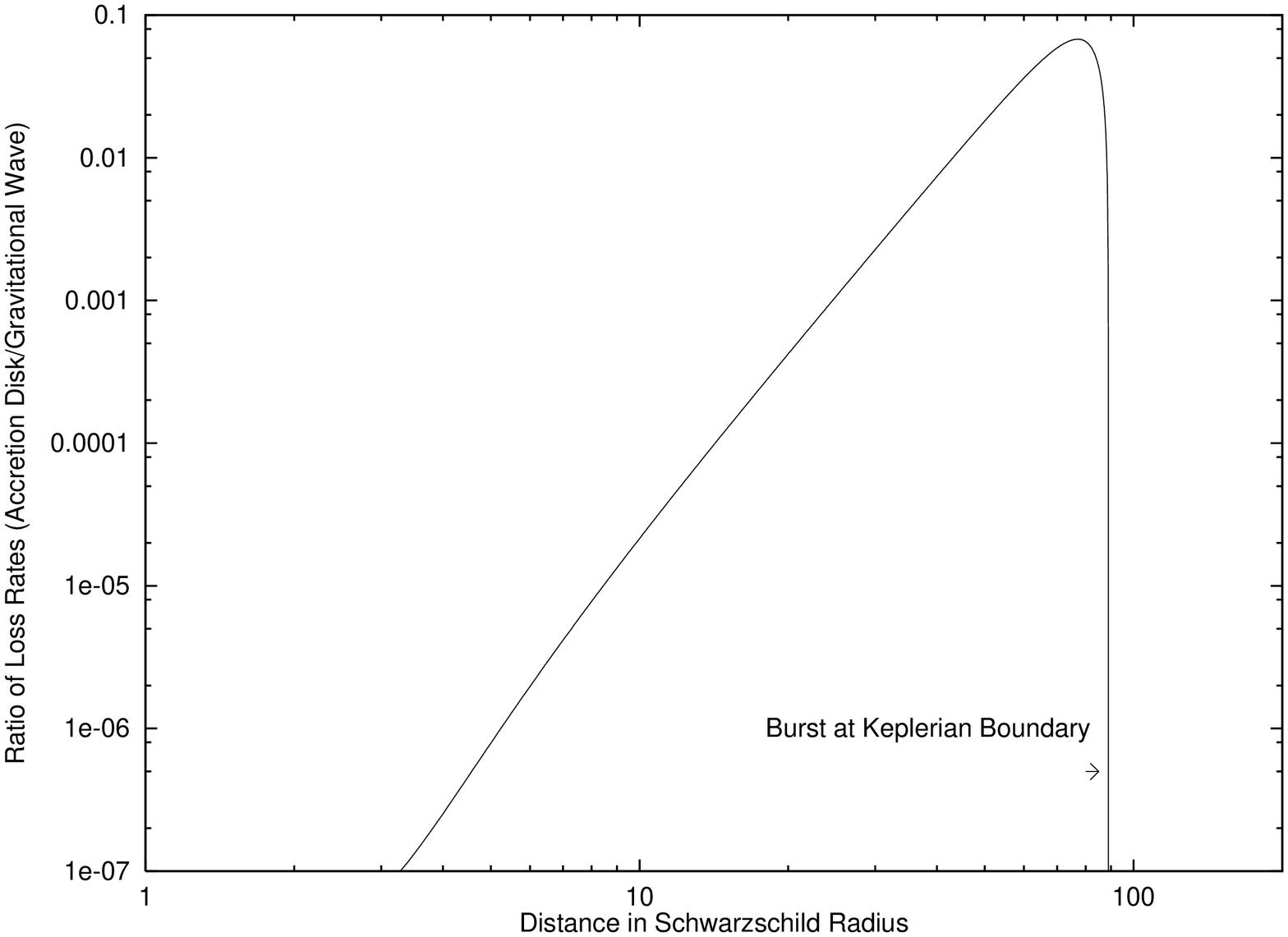,height=7.5truecm,width=7.5truecm,angle=270}
\noindent {\small {\bf Fig.1b}: Ratio of the rate of loss of 
angular momentum due to negative angular momentum accretion and 
the rate due to gravitational wave emission when the accretion 
rate is a  thousand times the Eddington rate.}

\psfig{figure=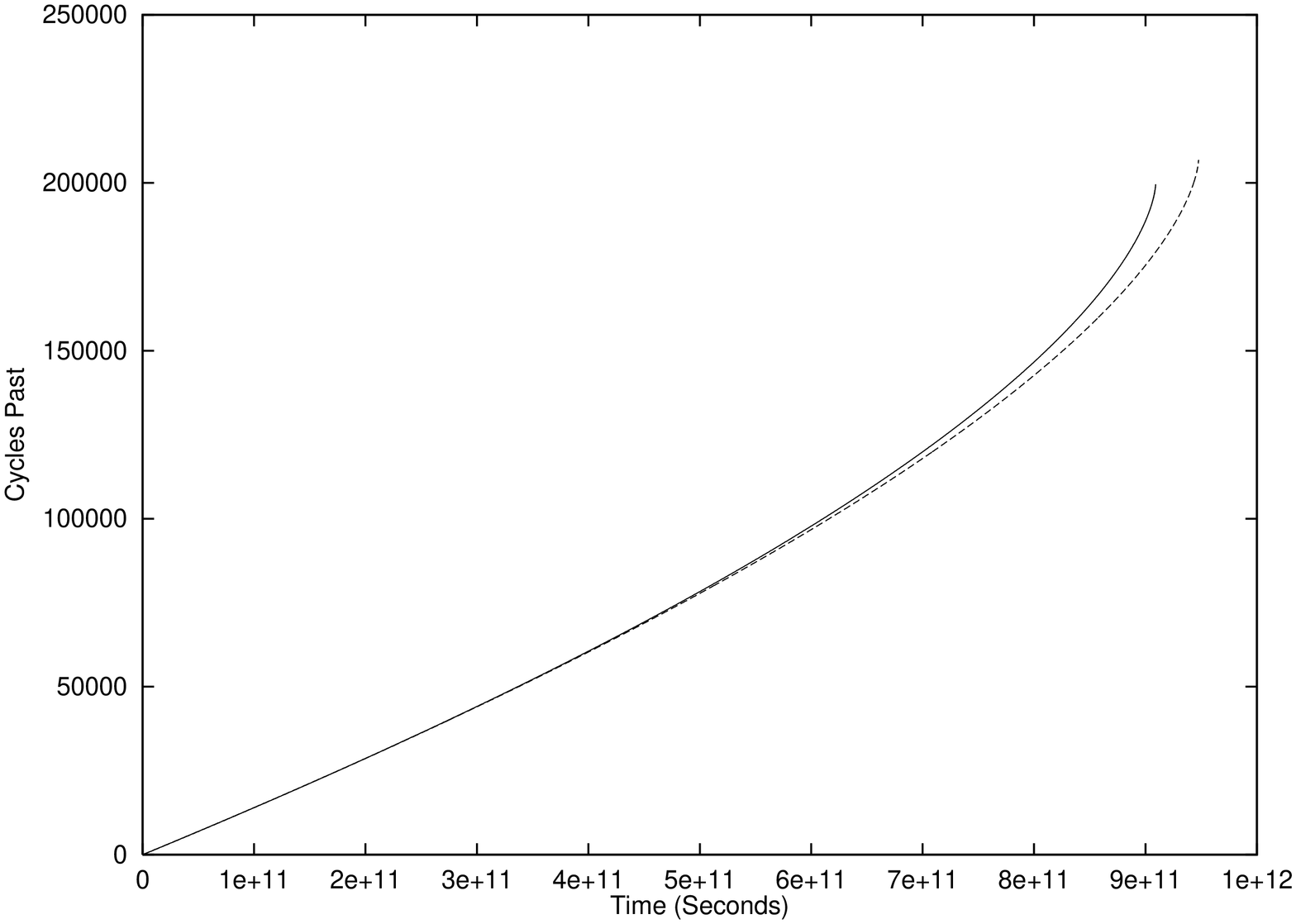,height=7.5truecm,width=7.5truecm,angle=270}
\noindent {\small {\bf Fig.1c}: Number of times companion orbits the
primary with (solid) or without (dashed) the disk taken into 
account. }

In some cases the accretion flow becomes {\it super-Keplerian}
before becoming sub-Keplerian near the horizon (Chakrabarti, 1996abc).
In these cases the low-mass companion will accrete positive angular
momentum from the disk, and in some extreme situation, this compensates
for the loss due to gravity waves. As a result, the companion orbit
will remain unchanged (Chakrabarti, 1993). Such systems, stability
of which has been verified by detailed  numerical simulations
(Molteni et al. 1994) would be steady sources of gravity waves
with constant frequency and amplitude.

\centerline{\bf References}

\noindent  Chakrabarti, S.K. 1993, ApJ, 411, 610\\
\noindent  Chakrabarti, S.K. 1996a, Physics Reports, 266, 229 \\
\noindent  Chakrabarti, S.K. 1996b, ApJ, 464, 664\\
\noindent  Chakrabarti, S.K. 1996c, Phys. Rev. D., 53, 290 \\
\noindent  Chakrabarti, S.K. \& Titarchuk, L.G., 1995, ApJ, 455, 623\\
\noindent  Molteni, D., Gerardi, G. \& Chakrabarti, S.K., ApJ, 436, 249\\
\noindent  Blanchet, L., et al., Phys. Rev. Lett., 1995, 74, 3515

\noindent To appear in the Proceedings of 6th Asia-Pacific Conference
(Journal of Korean Astronomical Society) August, 1996, Eds. H.M. Lee and S. Kim

\noindent Authors's address AFTER November 26th, 1996:\\

\noindent Prof. S.K. Chakrabarti\\
\noindent S.N. Bose National Center for Basic Sciences\\
\noindent JD Block, Sector -III, Salt Lake\\
\noindent Calcutta 700091, INDIA\\

\noindent e-mail: chakraba@bose.ernet.in  OR chakraba@tifrc2.tifr.res.in \\

\end{document}